\documentstyle[twocolumn]{jpsj}
\input{epsf.tex}
\newcommand{\bm}[1]{\mbox{\boldmath$#1$}}
\title{Role of Bond-Bond Interaction in the Extended Hubbard Chain}
\author{Masaaki {\sc Nakamura}, Kazuhito {\sc Itoh} and Norihiro {\sc
Muramoto}}
\inst{Institute for Solid State Physics,
University of Tokyo, Roppongi, Tokyo 106-8666}
\kword{site-off-diagonal interactions, g-ology, level-crossing approach,
strong-coupling theory, bond-charge-density wave, bond-spin-density
wave, exact solution}
\def\runauthor{Masaaki {\sc Nakamura}, Kazuhito {\sc Itoh}
and Norihiro {\sc Muramoto}}
\def\runtitle{Role of Bond-Bond Interaction in the Extended Hubbard Chain}
\recdate{}
\abst{
 Phase diagrams of the one-dimensional extended Hubbard model with a
 nearest neighbor interaction $V$ and site-off-diagonal (bond-charge $X$
 and bond-bond $W$) interactions at half-filling are obtained
 numerically, using the level-crossing approach for finite-size
 clusters.  This analysis is straightforward extension of the preceding
 work [cond-mat/9909277], where only the $X$ term is considered as a
 site-off-diagonal term.  Similarly to the $X$ term, the $W$ term
 enlarges bond-charge-density-wave (BCDW) and bond-spin-density-wave
 (BSDW) phases for $W<0$ and $W>0$ cases, respectively.  In the
 strong-coupling region, however, the $W$ term causes ferromagnetism for
 $W>0$, and shifts the boundary of the phase separation in the
 $U/t\ll-1$ region, which are different from the roles of the $X$ term.
 In addition, we discuss a parameter region where the BSDW state is the
 exact ground state, and show that the exactly solvable point appears
 just on the phase boundary between the BSDW and the ferromagnetic
 phases.}
\begin{document}
\sloppy
\maketitle 

\section{Introduction}
The Hubbard model has been considered to be one of the simplest models
to describe electrons in metals.\cite{Hubbard,Kanamori,Gutzwiller}
However, since this model is derived by neglecting elements of the
Coulomb integral except for the on-site term,\cite{Hubbard} it may be
unsufficient to describe phenomena where the neglected interactions play
essential roles.  The most dominant interaction next to the on-site term
is considered to be a nearest-neighbor density-density interaction $V$.
The Hubbard model including the $V$ term is usually called ``the
extended Hubbard model''. In addition, site-off-diagonal elements of the
Coulomb integral give bond-charge $X$ and bond-bond $W$ interactions.
Recently, effects of the site-off diagonal terms are pointed out in the
real materials.\cite{Ikeda-O} In this paper, we consider the role of
these site-off-diagonal interactions in one-dimensional (1D) cases,
where non-perturbative treatment is possible.

For this purpose, we consider the following generalized Hubbard model in
the half-filled band,
\begin{eqnarray}
 {\cal H}&=&-t\sum_{i}B_{i,i+1}
 +U\sum_i n_{i\uparrow}n_{i\downarrow}+V\sum_i n_{i}n_{i+1}\nonumber\\
&&+W\sum_{i}(B_{i,i+1})^2
+X\sum_{i,s}B_{i,i+1,s}(n_{i,-s}+n_{i+1,-s})\nonumber\\
&&+X'\sum_{i,s}B_{i,i+1,s}n_{i,-s}n_{i+1,-s},\label{eqn:tUVWX}
\end{eqnarray}
where
\begin{equation}
 B_{ijs}\equiv c^{\dag}_{is} c_{js} + c^{\dag}_{js} c_{is},\ \ \ 
 B_{ij}\equiv\sum_{s}B_{ijs}.
\end{equation}
The three body term in eq.~(\ref{eqn:tUVWX}) ($X'$ term) is not derived
from the generalized Coulomb integral. This term is justified, for
example, as an effective interaction in the three band
model.\cite{Simon-A} In fact, the $W$ term is interpreted as an
effective interaction in electron-phonon systems.\cite{Voit89}

Analysis of this model ($X'=0$) in the half-filled band was performed by
Campbell {\it et al.} ten years
ago.\cite{Campbell-G-L_lett,Campbell-G-L} They discussed the existence
of charge-density-wave (CDW), ``bond-order-wave (BOW)'' and
ferromagnetic (FM) phases, but they did not discuss all the phases in
this model. Especially, the nature of the ``BOW'' phase was left
ambiguous.

The property of this model can be clarified by the level-crossing
approach which is based on the conformal field theory and the
renormalization group. This method enables us to determine the phase
boundaries with high accuracy from numerical data of the finite-size
clusters.  Using this technique, one of the authors obtained the phase
diagram of the model (\ref{eqn:tUVWX}) for
$W=0$\cite{Nakamura99a,Nakamura99b}.  In this paper, we apply this
approach to the case $W\neq0$, and clarify the roles of the $W$ term.

The most important purpose of this paper is to clarify the nature of the
``BOW'' phase.  Japaridze first pointed out using the bosonization
technique that two types of ``BOW'' phases may appear in half-filled 1D
systems by the effect of the site-off-diagonal
terms.\cite{Japaridze,Japaridze-M,Japaridze-K} The two ``BOW'' phases
are bond-charge-density-wave (BCDW) and bond-spin-density-wave (BSDW)
phases. Their order parameters are given by
\begin{eqnarray}
 {\cal O}_{\rm BCDW}(j)&=&(-1)^j \sum_s  B_{j,j+1,s}\label{eqn:O_BCDW},\\
 {\cal O}_{\rm BSDW}(j)&=&(-1)^j \sum_s sB_{j,j+1,s}\label{eqn:O_BSDW}.
\end{eqnarray}
Recently, Itoh {\it et al.} discovered that the BSDW state gives the
exact ground state in some parameter region of a generalized Hubbard
model.\cite{Itoh-N-M} In this paper, by using the same argument, we
discuss the exactly solvable cases of the present model
(\ref{eqn:tUVWX}), and clarify the relation to the numerical results.

\section{Site-off-diagonal terms}\label{sec:model}
To begin with, we consider the role of the site-off-diagonal terms by
intuitive discussion. First, we consider roles of the bond-charge
interaction ($X,X'$) terms. Throughout this paper, we fix the parameters
as $X'=-2X$ for simplicity. In this case, the bond-charge interaction
terms are rewritten as
\begin{equation}
 {\cal H}_{X}=X\sum_{is}
 (c^{\dag}_{is} c_{i+1,s}+\mbox{H.c.})(n_{i,-s}-n_{i+1,-s})^2.
\label{eqn:X-term}
\end{equation}
This term conserves
SU(2)$\otimes$SU(2)/Z(2) symmetry of the Hubbard model under the
following canonical transformation,
\begin{equation}
c_{j\uparrow}\rightarrow c_{j\uparrow},\ \ \ 
c_{j\downarrow}\rightarrow (-1)^j c_{j\downarrow}^{\dag}.
\end{equation}
Hereafter, we call eq.~(\ref{eqn:X-term}) the $X$ term.  In a hopping
process of an electron with spin $s$, the hopping term and the $X$ term
gives $-t$ when the two sites have zero or two electrons with the
opposite spin $-s$, whereas they give $-t+X$ when the two sites have one
electron with spin $-s$. Therefore, we can expect that electron pairs
are stabilized when $X<0$.

Next, we consider roles of the bond-bond interaction ($W$) term. This
term can be rewritten as
\begin{equation}
 {\cal H}_{W}=
-4W\sum_i(\bm{S}_i\cdot\bm{S}_{i+1}+\bm{\eta}_i\cdot\bm{\eta}_{i+1}
-{\textstyle\frac{1}{4}}),
\end{equation}
where $\bm{S}_i$ is the usual spin operator, and $\bm{\eta}_i$ denotes
the $\eta$-pairing operator whose components are defined by
\begin{equation}
  \eta_i^{+}=(-1)^i c_{i\uparrow}^{\dag}c_{i\downarrow}^{\dag},\ \
  \eta_i^{-}=(-1)^i c_{i\downarrow}c_{i\uparrow},\ \
  \eta_i^{z}=\frac{1}{2}(n_i-1).
  \label{eqn:eta-pairing}
\end{equation}
Thus, the $W$ term includes the spin exchange and the pair
hopping. Since a pair hopping process gives an energy gain $2W$, one can
expect that the dimer state is enhanced for $W/t<0$. In addition, for
$W/t>0$, one can expect appearance of ferromagnetism for $U/t\gg 1$ and
a phase separation (ferromagnetism in the $\eta$-pairing space) for
$U/t\ll - 1$.

Apparently, the $W$ term also conserve the SU(2) $\otimes$ SU(2)/Z(2)
symmetry of the Hubbard model, so that the present model keeps this
symmetry when $V=0$. The SU(2) symmetry of the $\eta$-pairing is broken
when $V\neq 0$, and the entire symmetry is reduced to
U(1)$\otimes$SU(2).

\section{Weak-coupling region}\label{sec:weak}
The instability of the model in the weak-coupling limit can be discussed
based on the g-ology.\cite{Japaridze-K} The low-energy behavior of
eq.~(\ref{eqn:tUVWX}) at half-filling is described by two sine-Gordon
models for the charge and the spin sectors. Then, the phase boundaries
are determined in terms of the bare coupling constants.  According to
the results, the spin sector with SU(2) symmetry has a gap for
\begin{equation}
U<2V-\delta g,
\end{equation}
where $\delta g=4X/\pi+8W$.  On the other hand, the charge sector with
U(1) symmetry has a gap in the following region
\begin{eqnarray}
V&>&0,\ U<\delta g,\label{eqn:SU2-BKT}\\
U&>&-2V+\delta g,\ U>\delta g.
\end{eqnarray}
This region is given by two Berezinskii-Kosterlitz-Thouless (BKT)
transition lines.  Note that the BKT line given by $V=0$ reflects the
SU(2) symmetry of the $\eta$-paring.  In the charge-gap region, there is
a valley of the gap, and the gap vanishes on the Gaussian transition
line,
\begin{equation}
U=2V+\delta g,\ V<0.
\end{equation}
In the g-ology analysis, both $X$ and $W$ appear only through $\delta
g$, so that they play similar roles in the weak-coupling cases.  For
$\delta g=0$, CDW and spin-density-wave (SDW) phases appear in the
charge-gap region, while in the metallic region, two phases appear where
singlet or triplet superconducting (SS, TS) correlation is dominant.
The BCDW appears for $\delta g<0$ while the BSDW appears for $\delta
g>0$, between the CDW and the SDW phases.

Although the result of the g-ology is valid only in the weak-coupling
limit, the description by the sine-Gordon model does not break down in
the finite-coupling region, due to the concept of the Tomonaga-Luttinger
liquid.  In this case, the instabilities described by the sine-Gordon
model can be identified as level-crossings in the excitation spectra in
finite-size systems.  In this approach, logarithmic corrections are
canceled on the transition points, and the dependence of the system size
$L$ appears as ${\cal O}(1/L^2)$. This correction originates from the
deviation from the linearized dispersion relation of the
Tomonaga-Luttinger model.  Therefore, phase diagrams can be obtained
with high accuracy from the numerical data of the finite-size clusters.
The detail of the level-crossing approach is discussed in
ref.~\ref{Nakamura99b}.

\section{Strong-coupling region}\label{sec:strong}
The differences between the role of the $X$ and the $W$ terms are
clarified in the strong-coupling cases.  First, we consider the
transition between the phase-separated (PS) state and the SS phase, and
the one between the SS and the CDW states.  In the $U/t\ll-1$ region,
the model can be mapped onto the spin-$1/2$ XXZ spin chain using the
$\eta$-pairing operators defined in eq.~(\ref{eqn:eta-pairing}) and the
second-order perturbation theory\cite{Emery76,Fowler}:
\begin{equation}
 {\cal H}_{\rm XXZ}=J\sum_i\left[
\frac{1}{2}(\eta^{+}_i\eta^{-}_{i+1}+\eta^{-}_i\eta^{+}_{i+1})
+\Delta \left(\eta^{z}_i\eta^{z}_{i+1}-\frac{1}{4}\right)
\right],\label{eqn:XXZ}
\end{equation}
where
\begin{equation}
 J=\frac{4(t-X)^2}{|U|}-4W,\ \ J\Delta=J+4V.
\end{equation}
In the XXZ spin chain, $\Delta=1$ gives a BKT-type transition between
the XY (spin fluid) and the N\'{e}el states, and $\Delta=-1$ gives the
first-order transition between the XY and the ferromagnetic phases.  In
the present case, the PS, the SS and the CDW states correspond to the
ferromagnetic, the XY and the N\'{e}el states, respectively. Therefore,
$V=0$ gives the SS-CDW transition, and the asymptotic boundary of the
SS-PS transition is given by
\begin{equation}
 V=-\frac{2(1-\xi)^2t^2}{|U|}+2W,\label{eqn:ps1}
\end{equation}
where $\xi\equiv X/t$.  Note that this result is valid only for $W/t\leq
0$, because the SS-PS boundary do not intersect the SS-CDW line ($V=0$)
due to the property of the XXZ spin chain.

Next, we calculate the energies of the CDW and the SDW states by
extending van Dongen's perturbative expansion.\cite{Dongen} Since the
CDW state is described as a ``N\'{e}el-ordered phase'' ($\Delta>1$) in
the $\eta$-pairing space, the energy is given by
\begin{eqnarray}
 \frac{E_{\rm CDW}}{L}&=&
\frac{U}{2}\label{eqn:E_CDW}
+4|W|\,\varepsilon_{\rm XXZ}(|V/W-1|)
-\frac{2(1-\xi)^2t^2}{(3v-1)U}\\
&&\hspace{-1cm}
 {}+\frac{(1-\xi)^2\left[(36v^2-5v-1)(1-\xi)^2-8(3v-1)v\right]t^4}
{v(3v-1)^3(4v-1)U^3},\nonumber
\end{eqnarray}
where $v\equiv V/U$.  $\varepsilon_{\rm XXZ}(\Delta)$ is the energy
density of the XXZ spin chain (\ref{eqn:XXZ}) with $J=1$, in the
N\'{e}el-ordered region ($1<\Delta$),
\begin{equation}
 \varepsilon_{\rm XXZ}(\cosh\theta)=
-\sum_{n=-\infty}^{\infty}
\frac{\sinh\theta}{1+{\rm e}^{-2|n|\theta}},
\end{equation}
where $\cosh\theta\equiv\Delta$. On the other hand, the perturbative
expansion for the SDW state can be performed up to the fourth order,
using the Bethe-ansatz result for the spin-$1/2$ Heisenberg chain as
\begin{eqnarray}
 \frac{E_{\rm SDW}}{L}&=&
 V-\left[\frac{(1-\xi)^2t^2}{(1-v)U}-W\right]4\ln 2
 \label{eqn:E_SDW}\\
&&
 {}+9\zeta(3)\frac{(1-\xi)^2\left[2(1-\xi)^2-1+v\right]t^4}
  {(1-v)^3U^3}.
\nonumber
\end{eqnarray}
The expansion for the SDW state is valid only for $W\leq 0$.

The energies for the PS and the FM states are not affected by the
perturbation in the thermodynamic limit.  In the PS state, the system
separates into two domains: doubly occupied sites and a
vacuum.\cite{Lin-H} Then, the energy in the thermodynamic limit is given
by
\begin{equation}
  E_{\rm PS}=\frac{U+4V}{2}L.\label{eqn:E_PS}
\end{equation}
On the other hand, the energy of the FM state is given by
\begin{equation}
  E_{\rm FM}=VL.\label{eqn:E_FM} 
\end{equation}

Using the energies obtained above, we determine asymptotic forms of the
four first-order transition lines by equating the energies as
\begin{eqnarray}
 E_{\rm CDW}&=&E_{\rm SDW},\label{eqn:CDW-SDW}\\
 E_{\rm CDW}&=&E_{\rm FM},\label{eqn:CDW-FM}\\
 E_{\rm SDW}&=& E_{\rm PS},\label{eqn:ps2}\\
 E_{\rm FM}&=&E_{\rm PS}.\label{eqn:FM-PS}
\end{eqnarray}
In the $U,V\rightarrow\infty$ limit, the CDW-SDW phase boundary is
$V-U/2=2|W|(2\ln 2-1)$ for $W\leq 0$, while the CDW-FM phase boundary is
$V-U/2=2W$ for $W>0$. The FM-PS boundary is fixed at $U=-2V$.

In the present strong-coupling theory, $X$ appears in the every order of
$t^{2n}/U^{2n-1}$ in the perturbation series, while $W$ appears only in
the zeroth order. Therefore, the $W$ term plays roles to shift the
boundary lines of the phase separation, the CDW-SDW and the CDW-FM
transitions.

\section{Exactly solvable case}\label{sec:exact}

We discuss the exact ground state of the BSDW state in the present
model, using the argument in ref.~\ref{Itoh-N-M}, where the three body
term (the $X'$ term) is not included.  The explicit wave functions for
the BCDW and the BSDW states are given by products of ``electron-hole
dimers'' as,
\begin{eqnarray}
 |{\rm BCDW}\rangle_s&=&
\prod_{i=1}^{L/2}[2i-1,2i]_{s}[2i-1,2i]_{-s}|0\rangle,
\label{eqn:BCDW}\\
 |{\rm BSDW}\rangle_s&=&
 \prod_{i=1}^{L/2}[2i-1,2i]_{s}[2i,2i+1]_{-s}|0\rangle,
\label{eqn:BSDW}
\end{eqnarray}
where $[i,j]_s\equiv
\frac{1}{\sqrt{2}}(c^{\dag}_{is}+c^{\dag}_{js})$. These ``electron-hole
dimers'' can be interpreted as electrons located on the bonds.

Now, we consider a local operator $h_{ijs}$ for spin-$s$ which satisfies
$h_{ijs}[i,j]_s|0\rangle=0$, and $h_{ijs}|{\rm otherwise}\rangle\geq 0$
for the other configurations. For example,
\begin{subequations}
\begin{eqnarray}
 h_{ijs}^{a}&=&-B_{ijs}+n_{is}+n_{js},\\
 h_{ijs}^{b}&=&-B_{ijs}-n_{is}-n_{js}+2,\\
 h_{ijs}^{c}&=&-B_{ijs}+1,\\
 h_{ijs}^{d}&=&-B_{ijs}+B_{ijs}^2.
\end{eqnarray}
\end{subequations}
If the total Hamiltonian is written in the following form using the
local operators as
\begin{equation}
 {\cal H}=\sum_{ij}\lambda_j
  h_{i,i+1,\uparrow}^{a_j}h_{i,i+1,\downarrow}^{b_j},
  \ \ \ \lambda_j>0,
 \label{eqn:Hamiltonian_prod0}
\end{equation}
then ${\cal H}|{\rm BSDW}\rangle_s=0$.  Therefore, the BSDW state is the
exact ground state of this Hamiltonian.\cite{Itoh99a} The present model
(\ref{eqn:tUVWX}) can be written using the above local operators as
follows
\begin{eqnarray}
 {\cal H}-E&=&
 t\sum_{i}\left[{\textstyle\frac{1-\lambda}{2}}
(h_{i,i+1,\uparrow}^{a} h_{i,i+1,\downarrow}^{a}
+h_{i,i+1,\uparrow}^{b} h_{i,i+1,\downarrow}^{b})\right.\nonumber\\
&&\left.+\lambda
(h_{i,i+1,\uparrow}^{c} h_{i,i+1,\downarrow}^{d}
+h_{i,i+1,\uparrow}^{d} h_{i,i+1,\downarrow}^{c})\right],
 \label{eqn:Hamiltonian_prod}
\end{eqnarray}
where $0\leq\lambda\leq 1$, and the parameters and the ground-state
energy of the model are
\begin{subequations}\label{eqn:pram_region}
\begin{eqnarray}
 U&=&2V=2(1-\lambda)t,\\
 X&=&-X'/2=-\lambda t,\\
 W&=&(1+\lambda)t/2,\\
 E&=&(1-\lambda)tL.
\end{eqnarray}
\end{subequations}

The Hamiltonian (\ref{eqn:tUVWX}) has SU(2) symmetry under global
rotation in the spin space. In other words, the operator for the total
spin $S^{\alpha}_{\rm tot}\equiv\sum_i S^{\alpha}_i$ commutes with the
Hamiltonian, so that $(S^{+}_{\rm tot})^{n}|{\rm BSDW}\rangle_s$ with
$n\leq L/2$ also gives the ground state.  Therefore, the fully
ferromagnetic (FM) state ($n=L/2$),
\begin{equation}
 |{\rm FM}\rangle=
 \prod_{i=1}^{L}c^{\dag}_{is}|0\rangle,
\end{equation}
also degenerates with the BSDW state. This large degeneracy is
consistent with the result of the bosonization theory where a BSDW is
described as bond-located spins with gapless
excitations.\cite{Japaridze,Nakamura99b}

In finite-size systems, there is no spontaneous breaking of the
translational symmetry, so that the actual wave functions of the BSDW
states are reconfigured by linear combinations of eq.~(\ref{eqn:BSDW})
for $s=\uparrow$ and $\downarrow$ as
\begin{equation}
  |{\rm BSDW}\rangle_{\pm}=
  |{\rm BSDW}\rangle_{\uparrow}\pm |{\rm BSDW}\rangle_{\downarrow}.
\end{equation}
These two states appear at two different wave numbers $k=0,\pi$.  When
the parameters of the model shifts from the exactly solvable case toward
the spin-gapless region, the $k=0$ mode remains the ground state, while
the $k=\pi$ mode gives the lowest excited state.

We can also construct a model which has a BCDW state as an exact ground
state, if the local operators in eq.~(\ref{eqn:Hamiltonian_prod0}) is
replaced as $h_{i,i+1,\uparrow}^{a_j}\rightarrow
h_{i-1,i,\uparrow}^{a_j}$. In this case, the obtained Hamiltonian
includes interactions among three sites.

\section{Numerical results}
\begin{figure}[h]
\begin{center}
\noindent
\epsfxsize=9cm \leavevmode \epsfbox{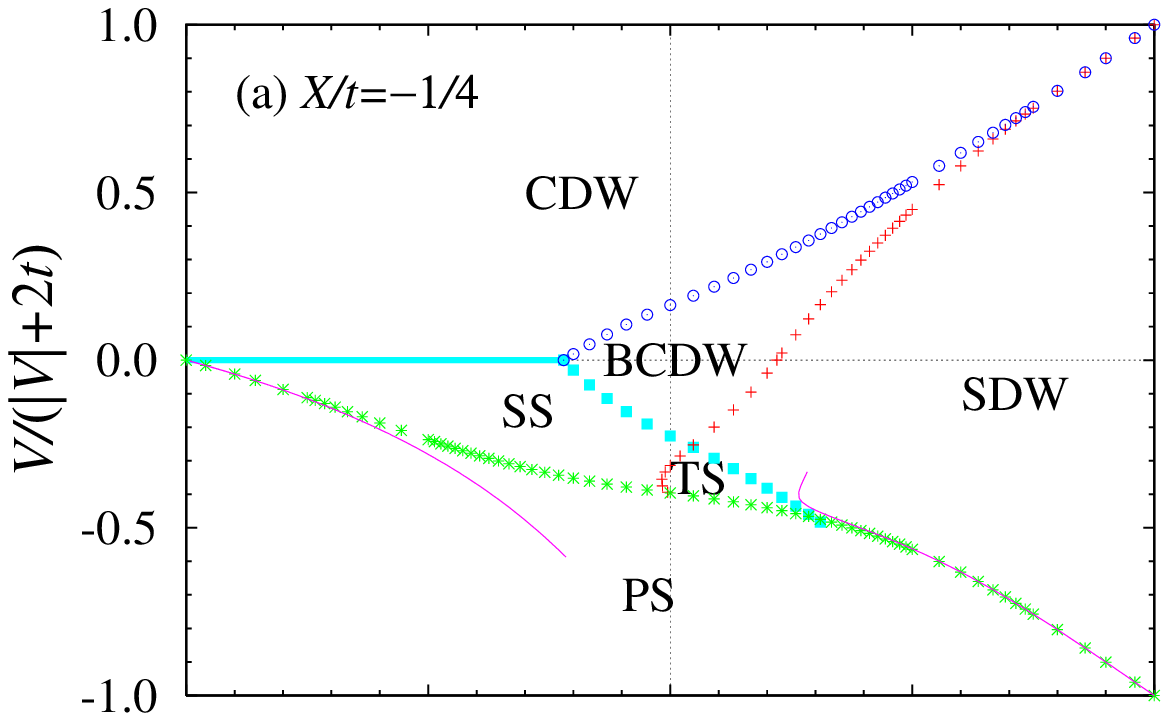}
\epsfxsize=8cm \leavevmode \epsfbox{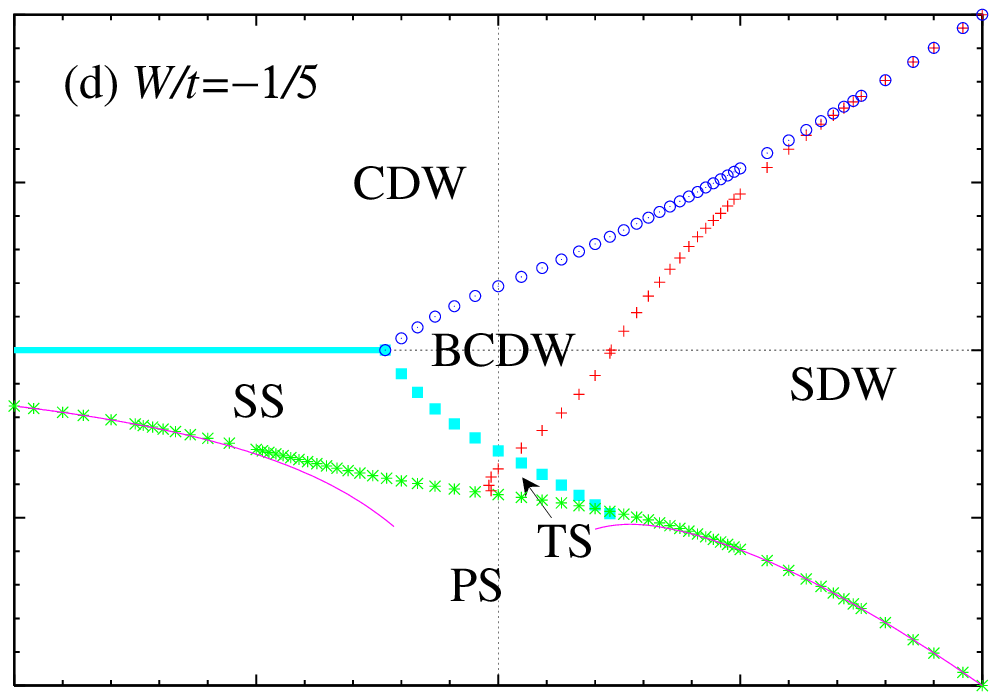}\\
\epsfxsize=9cm \leavevmode \epsfbox{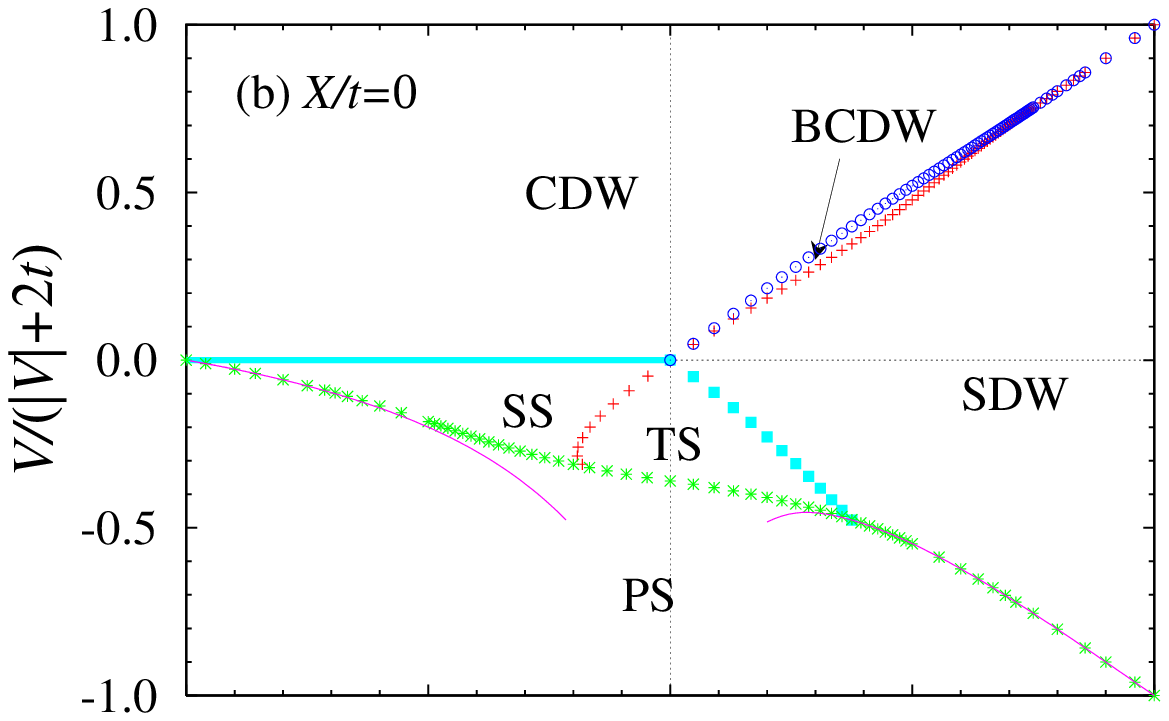}
\epsfxsize=8cm \leavevmode \epsfbox{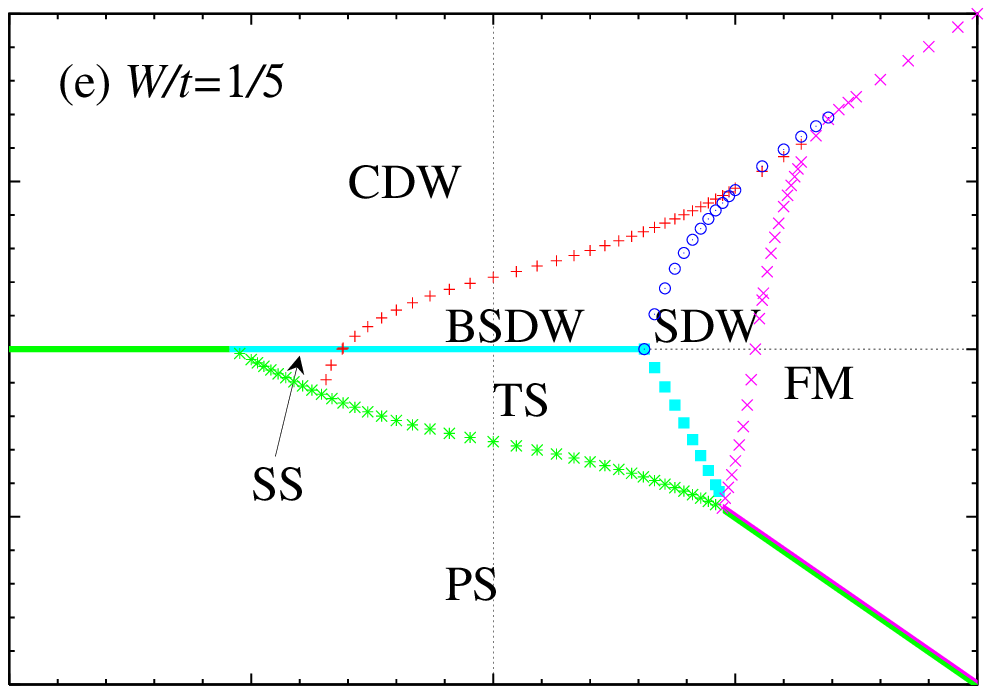}\\
\epsfxsize=9cm \leavevmode \epsfbox{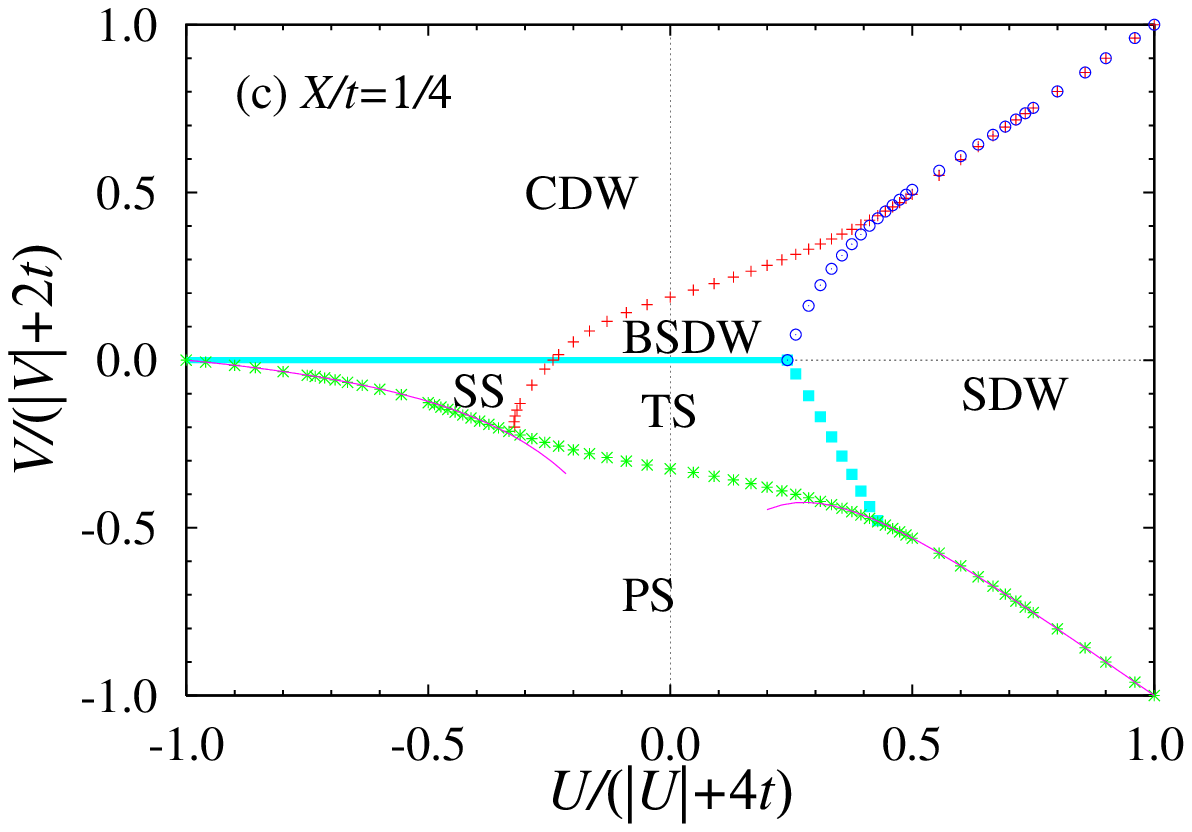}
\epsfxsize=8cm \leavevmode \epsfbox{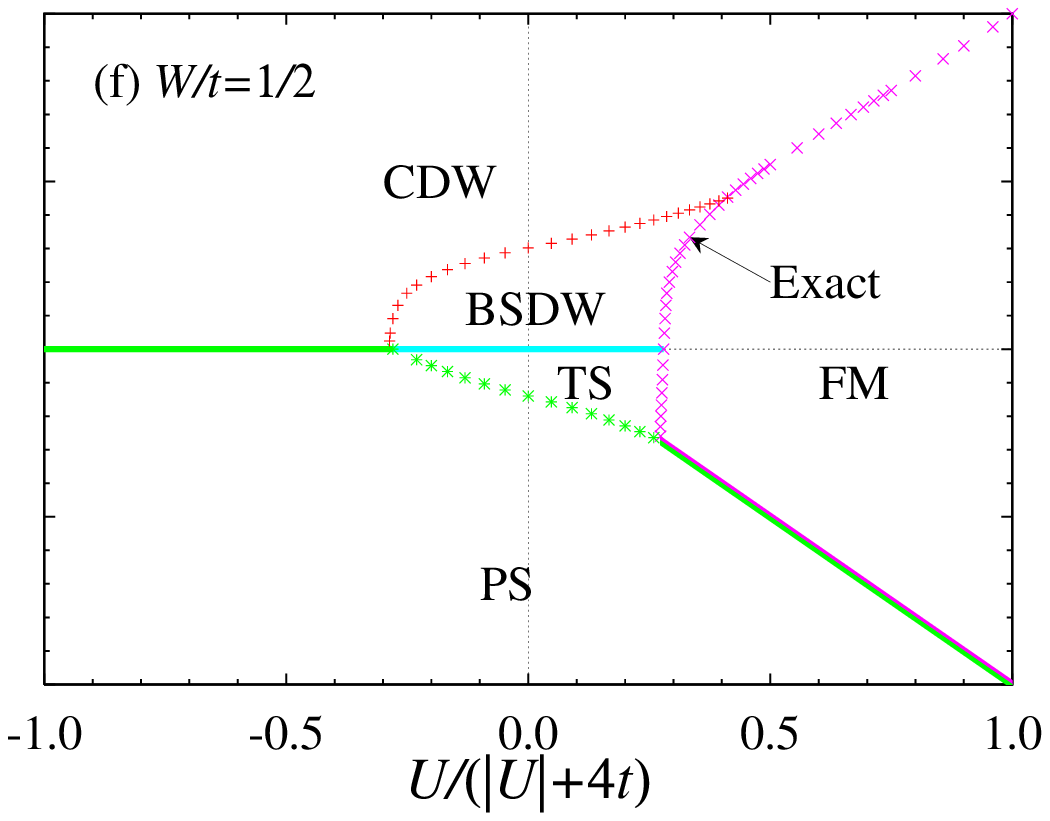}
\end{center}  
\caption{Phase diagram of the 1D extended Hubbard model with
 site-off-diagonal interactions determined by the data of the $L=12$
 system at half-filling.  [CDW (SDW): charge- (spin-) density wave, BCDW
 (BSDW): bond-CDW (bond-SDW), SS (TS): singlet (triplet) superconducting
 phase, PS: phase-separated state, FM: ferromagnetic phase] The
 asymptotic phase boundaries for the PS are given by
 eqs.~(\protect{\ref{eqn:ps1}}) and (\protect{\ref{eqn:ps2}}). (a)-(c)
 are taken from ref. \ref{Nakamura99b} for comparison.}  \label{fig:PD1}
\end{figure}

Using the level-crossing approach as in the same way of
ref.~\ref{Nakamura99b}, we determine the phase diagrams of
eq.~(\ref{eqn:tUVWX}) in the $U$-$V$ plain for $W\neq 0$, $X=X'=0$ by
analyzing the numerical data of the exact diagonalization for $L=12$
systems. We determine first-order transition lines for the FM and the PS
states by comparing the ground-state energies with eqs.~(\ref{eqn:E_FM})
and (\ref{eqn:E_PS}).  Then, we obtain the results as shown in
Figs.~\ref{fig:PD1}(d)-(f). For comparison, we also show the phase
diagrams for $W=0$, $X=-X'/2\neq 0$ in Figs.~\ref{fig:PD1}(a)-(c) which
are taken from ref.~\ref{Nakamura99b}.

For $W<0$, the BCDW phase appears between CDW and SDW phases, while for
$W>0$, a BSDW phase appears instead of the BCDW phase. This property is
similar to the effects caused by the $X$ term. The result is consistent
with the estimation that a dimer (BCDW) state is stabilized for $X,W<0$
as was discussed in Sec.~\ref{sec:model}.  In ref.~\ref{Campbell-G-L},
existence of the ``BOW'' state is discussed for $X=X'=0, W>0$, so that
this ``BOW'' phase corresponds to the BSDW state in the present results.

The differences between the roles of the $X$ and the $W$ terms are
clarified in the strong-coupling region.  One is a property of the PS
boundary in the $U/t\ll -1$ region.  For $W<0$, the PS boundary shifts
toward the negative-$V$ side, and flows into $(U,V)=(-\infty,-2|W|)$. On
the other hand, for $W>0$, the boundary shifts toward the positive-$V$
side, but it does not cross the $V=0$ line.  Therefore, a direct CDW-PS
transition takes place at $V=0$.

The other difference is appearance of a FM state for $W>0$.  For
$W/t=1/5$, a tetracritical point appears among PS, TS, SDW and FM
states.  As $W$ is increased, the SDW region and the Gaussian line in
the charge part are swallowed by the FM region, and a direct BSDW-FM
transition takes place.

\begin{figure}[h]
\noindent
\epsfxsize=3.4in \leavevmode \epsfbox{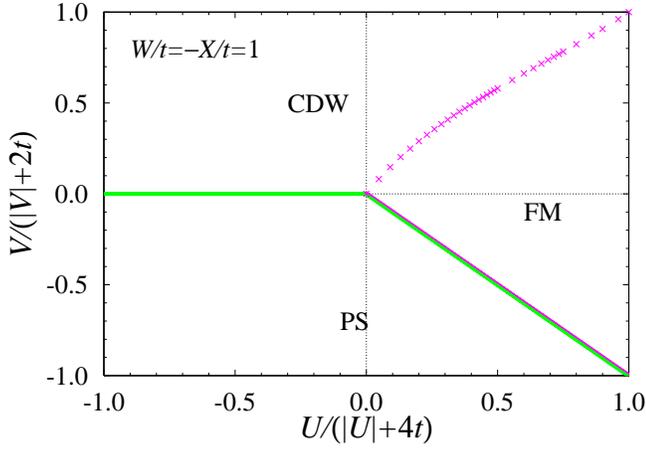}
\caption{Phase diagram of the 1D extended Hubbard model with
site-off-diagonal interactions ($W/t=-X/t=1$) determined by the data of
the $L=12$ system at half-filling.}  \label{fig:PD2}
\end{figure}

For $W/t=1/2$, an exactly solvable point ($\lambda=0$) appears at
$(U/t,V/t)=(2,1)$.  This point is just on the boundary between the BSDW
and the FM phases as was discussed in Sec.~\ref{sec:exact}. The exactly
solvable point moves on the $U=2V$ line as $\lambda$ is increased.  When
$W/t=-X/t=1$ ($\lambda=1$), exactly solvable point appears at
$(U/t,V/t)=(0,0)$. In this case, CDW, PS, FM and BSDW phases degenerates
on this point.  The other regions of the phase diagram are occupied only
by CDW, PS and FM states (see Fig.~\ref{fig:PD2}).

\begin{figure}[h]
\noindent
\epsfxsize=1.7in \leavevmode \epsfbox{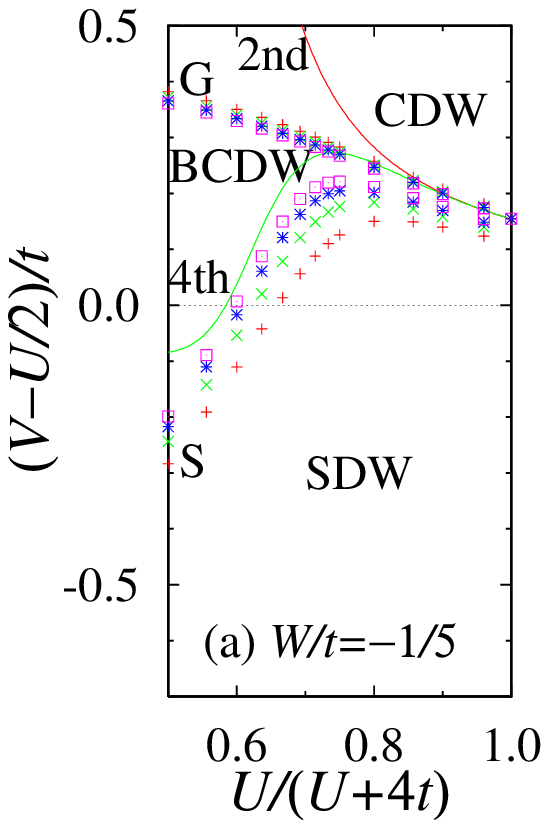}
\epsfxsize=1.7in \leavevmode \epsfbox{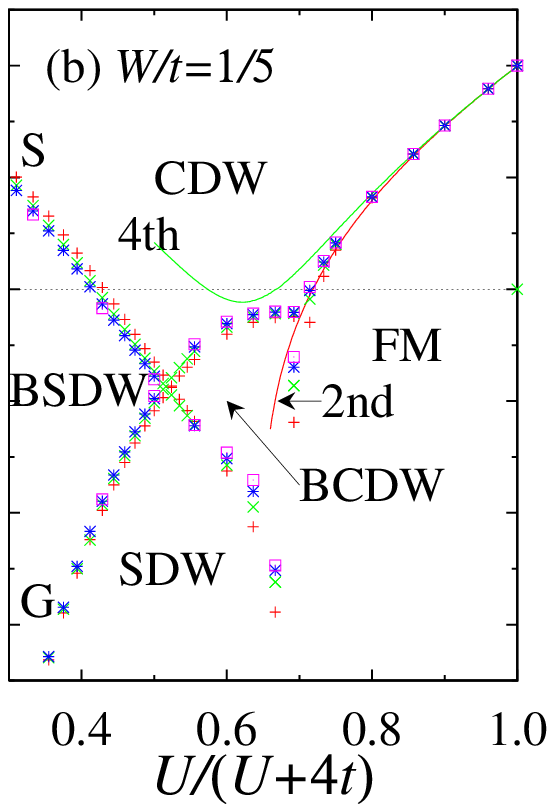}\\
\epsfxsize=3.4in \leavevmode \epsfbox{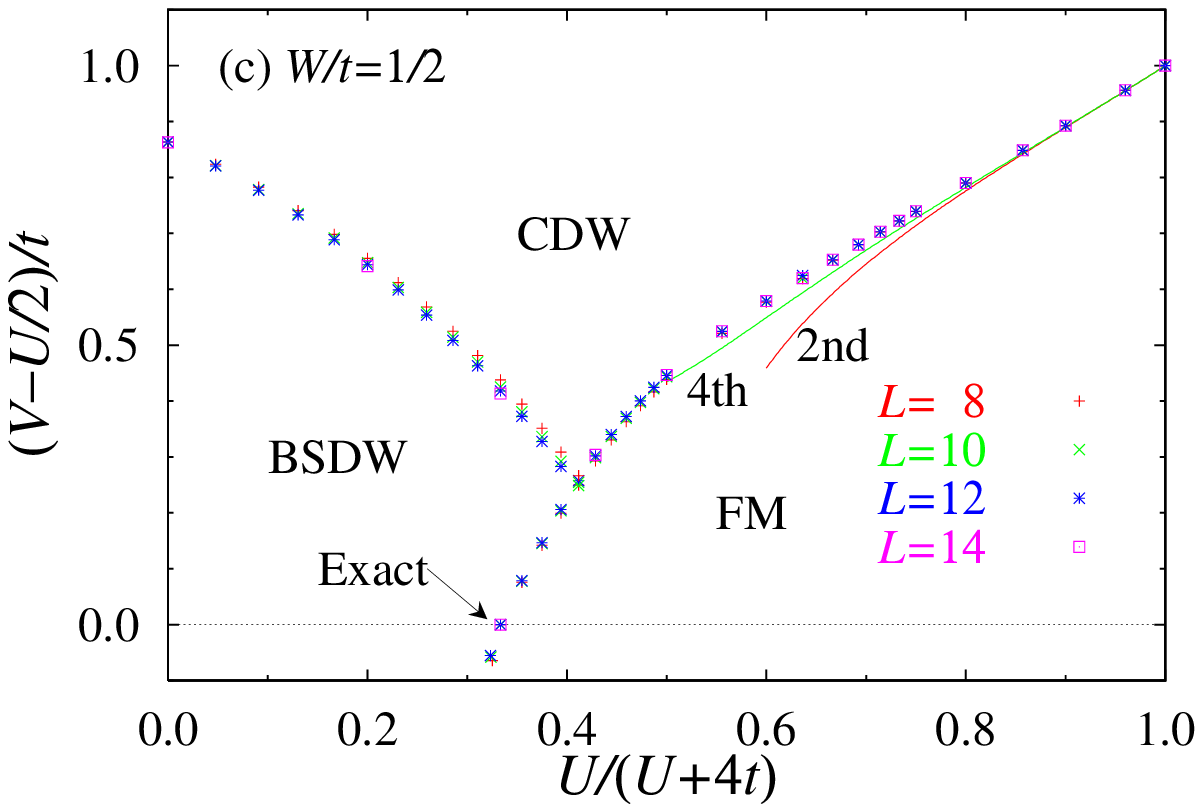}
\caption{Phase boundaries [(G) Gaussian transition for the charge
 sector, (S) spin-gap transition] near the $U=2V$ line for (a)
 $W/t=-1/5$, (b) $W/t=1/5$ and (c) $W/t=1/2$, determined in $L=8$-$14$
 systems.  The strong-coupling results [eqs.~(\ref{eqn:CDW-SDW}) and
 (\ref{eqn:CDW-FM})] agrees with the Gaussian line of the charge sector
 in (a), and the CDW-FM boundaries in (b) and (c), respectively.}
 \label{fig:diffs}
\end{figure}

In order to check the validity of the result in the strong-coupling
region, we show in Fig.~\ref{fig:PD1} (d), asymptotic forms of the PS
boundaries given by eqs.~(\ref{eqn:ps1}) and
(\ref{eqn:ps2}).\cite{Lin-H} They well agree with the numerical result.
In Fig.~\ref{fig:diffs}, we show the phase boundaries near the $U=2V$
line for $L=8$-$14$ systems, and asymptotic boundaries given by the
perturbation theory up to the second and the fourth orders. In
Fig.~\ref{fig:diffs} (a), the CDW-SDW transition line given by
eq.~(\ref{eqn:CDW-SDW}) agrees with the Gaussian line in the charge
part.  Although the size dependence of the the spin-gap phase boundary
is large, it also approaches to the same line in the strong-coupling
regime as the system size is increased.  In Figs.~\ref{fig:diffs} (b)
and (c), eq.~(\ref{eqn:CDW-FM}) agrees with the CDW-FM transition lines. 
In Fig.~\ref{fig:diffs} (b), the BCDW phase appears in the very narrow
region, and a tetracritical point appears among the BSDW, the BCDW, the
SDW and the CDW phases.

\section{Summary}
We have studied the phase diagram of the extended Hubbard chain
including the bond-charge $X$ and the bond-bond $W$ interactions, using
the level-crossing approach. We have clarified the similarity of the
roles of the $X$ and the $W$ terms in the weak-coupling case (appearance
of BCDW and BSDW phases), and difference in the strong-coupling case
(appearance of a FM state, properties of the PS boundary).  We have also
shown the exact BSDW state appears just on the phase boundary between
the BSDW and the ferromagnetic regions.

\section{Acknowledgment}
The authors are grateful to M. Ogata and M. Takahashi for useful
discussions.  The computation in this work was partly done using the
facilities of the Supercomputer Center, Institute for Solid State
Physics, University of Tokyo.

\pagebreak

\onecolumn

\pagebreak

\twocolumn

\pagebreak


\end{document}